\documentclass[article,twocolumn,showpacs]{revtex4}
\usepackage{amssymb,amsmath,amsfonts}
\usepackage{graphicx}

\begin{document}

\title{Can surface cracks and unipolar arcs explain breakdown and gradient limits?}

\author{ Z. Insepov, J. Norem$^*$}
\affiliation{Argonne National Laboratory, Argonne, IL 60439, USA}
\email{norem@anl.gov} 
 
\date{\today}

\begin{abstract}
We argue that the physics of unipolar arcs and surface cracks can help understand rf breakdown, and vacuum arc data.  We outline a model of the basic mechanisms involved in breakdown and explore how the physics of unipolar arcs and cracks can simplify the picture of breakdown and gradient limits in accelerators, tokamaks as well as laser ablation, micrometeorites and other applications.  Cracks are commonly seen in SEM images of arc damage and they are produced as the liquid metal cools, they produce the required field enhancements to explain field emission data data and can produce fractures that would trigger breakdown events.  Unipolar arcs can produce currents sufficient to short out rf structures, should cause the sort of damage seen in SEM images, should be unstable and possibly self-quenching as seen in optical fluctuations and surface damage.
\end{abstract}

\pacs{29.20.-c, 52.80.Vp}

\maketitle

\section{Introduction}
In this paper we explore how well arcing can be explained by the properties of surface cracks and unipolar arcs.  The question of explaining breakdown, arcing and gradient limits presents a unique problem, since there are over 100 years of reliable published data on an enormous variety of phenomena that seem related, but no simple explanation has been adopted that can easily be applied to clarify or predict the overall physics \cite{earhart,Anders,laurant,Juttner,mesyats}.  While it is always possible to introduce a variety specific mechanisms that can be narrowly applied, these may not be generally useful to explain or predict specific results.   We find that the mechanisms of unipolar arc physics, combined with surface cracking, can explain a significant fraction of the data, and a further analysis seems useful, however these mechanisms are not mentioned in most of the literature on arcing. Although most of our examples are from rf breakdown, the conclusions should have wider applicability.   The ultimate test of a model is whether the ideas are simple, complete and general enough to be useful.  This paper is an outline of these ideas.

\begin{figure}[htb]
    \centering
    \includegraphics*[width=88mm]{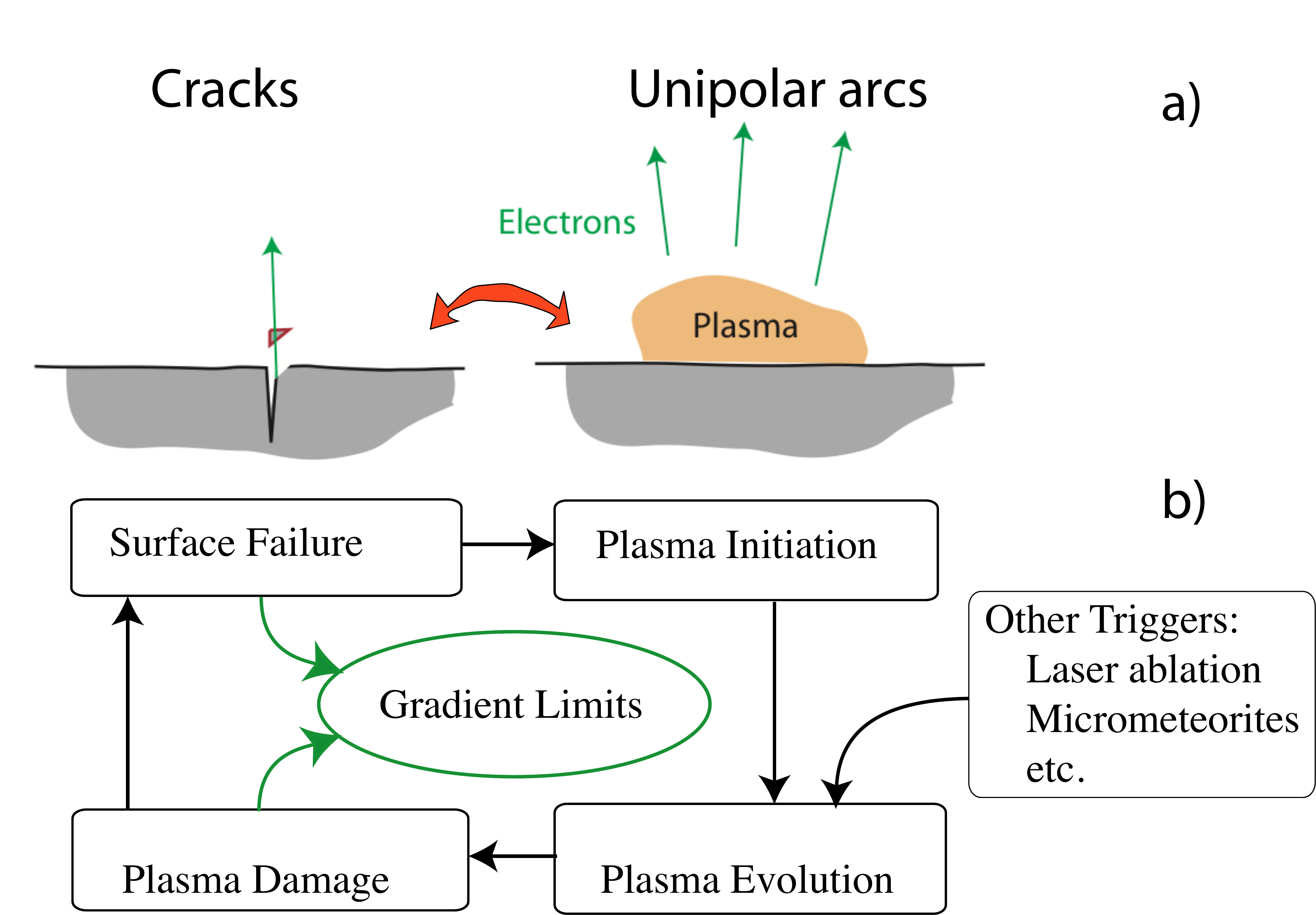}
    \caption{ The arc process is controlled by: 1) fracture of high field areas at crack junctions and, 2) the evolution of the unipolar arc driven by sheath parameters.}
\end{figure}

Our picture of arcs is summarized in Fig. 1 \cite{Noremrf2011,InsepovNorem11}.  We argue that two processes seem to control arcing:  1) the formation and fracture of cracks and small structures, and, 2) the evolution and properties of unipolar arcs .  Theoretically, we divide the arcing process itself into four elements:  1) mechanical failure of the surface, producing fragments, 2) initial ionization of the fragments by field emission (FE) currents, 3) exponential density growth of the unipolar arc to some equilibrium state, and, 4) surface damage produced by the arc.  The unipolar arcs act as virtual cathodes and produce currents that short the rf cavity or other high gradient structure.  

The study of these phenomena has been complicated by the speed and unpredictability of the arcs, as well as the large dynamic range of the experimental parameters and the numerical complications involved in simulations, where the densities involved seem to exceed the applicability of  the Particle-in-Cell (PIC) codes used for most plasma calculations.   The difficulties involved in accurately modeling plasma / surface interactions for very dense plasmas are a significant limitation on modeling \cite{Noremrf2011}.
  
Numerical modeling of the initiation of the arc using a Particle-in-Cell (PIC) has been described in a number of papers \cite{Noremrf2011,InsepovNorem11}.  Once an arc starts, the surface electric field and field emission increase, increasing ionization of neutrals, causing an increase in the plasma density.  This density increase decreases the Debye length and causes an increase in the surface electric field, ultimately producing an exponential increase in both the electric field and density, with time.  PIC simulations of the unipolar arc model for vacuum arcs relevant to rf cavity breakdown show that the density of plasma formed above the field emitting asperities can be as high as $10^{26}\ \mathrm{m}^{-3}$. The temperature of such plasma is low, in the range of $1-10$~eV. 

While we find that the basic mechanisms can be described simply and some results can be evaluated easily, however obtaining more precise results using numerical modeling is complicated by the multidisciplinary nature of the problems.  We will describe some of the basic mechanisms, simple results and more difficult calculations.  We show how these arguments apply to questions about field emission, breakdown, nonideal plasmas, plasma instabilities and quenching, arc suppression, gradient limits, frequency dependence, magnetic field effects, etc..

\section{Elements}
Since the literature on both surface cracking and unipolar arcs in this context is somewhat limited we review the the relevant physics of these phenomena.

\subsection{Arc evolution}
We have described how arc evolution can take place in rf structures and other environments, see Fig 1b.  We assume that the overall process can be divided into four more or less independent stages, surface failure, initial ionization, plasma evolution and surface damage.  If a dense plasma is created on the surface, by laser ablation for example, the further evolution of this plasma should be expected to be be similar to that of an rf plasma. Surface failure due to Maxwell stresses has been modeled by means of Molecular Dynamics (MD) and the initial ionization stage and  later plasma evolution by means of a PIC code.  The validity of the results of the PIC code is limited in the case of high density, non-ideal plasmas, but calculations of nonideal sheath plasmas have been done with MD codes.  We assume the limiting gradient for any system will be determined by a combination of the surface damage, which determines the local field enhancements,
\[E_{local}=\beta E_{average} \]
and the surface failure mechanism.    

\subsection{Unipolar Arcs}
Unipolar arcs were first described by Robson and Thonemann in 1959 as an explanation for the existence of isolated cathode spots on metal surfaces immersed in the plasma of a gas discharge \cite{Robson}.  Unipolar arc phenomena received extensive study and analysis in the 1970's and 1980's as the primary mechanism that determined the impurity content of limiter tokamaks.  Schwirzke and others described both experimental and theoretical work with these arcs \cite{Schwirzke91}.  As more tokamaks were built with divertors, however this mechanism seemed to become less relevant, although that may be changing as the physics of the ITER tokamak is better understood \cite{ITER}.   The most recent and thorough study of unipolar arcs is being done by Kajita, who uses laser ablation to produce a plasma on a metallic surface that starts the unipolar arc phenomenon \cite{kajita}.  

Unipolar arcs can than travel freely on the surface or be guided by a magnetic field in the characteristic retrograde motion that has been identified in many experiments.  The high plasma densities are associated with a large plasma pressure which should be responsible for particulate production.  We have found that the interface between the plasma and the liquid surface can become turbulent due due to the high plasma pressure in a dense arc and the scale of the turbulence is a function of the plasma pressure.  Some of the parameters of the unipolar arc plasma can be experimentally estimated from the dimensions of the damage produced and measurements from SEM images imply the density is very high.

\subsection{Cracks}
Arrays of cracks are seen in many SEM images of arc damage.  We believe these cracks are the result of the cooling of the melted surface that takes place in two stages; first cooling from high temperatures to the solidification point of the metal, followed by cooling from the melting point to room temperature, where the solid contracts by an amount $\Delta x = x \alpha \Delta T$, where $T$ is the temperature, $x$ represents the dimensions of the damage and $\alpha$ is the coefficient of linear expansion. 

\begin{figure}[htb]
    \centering
    \includegraphics*[width=88mm]{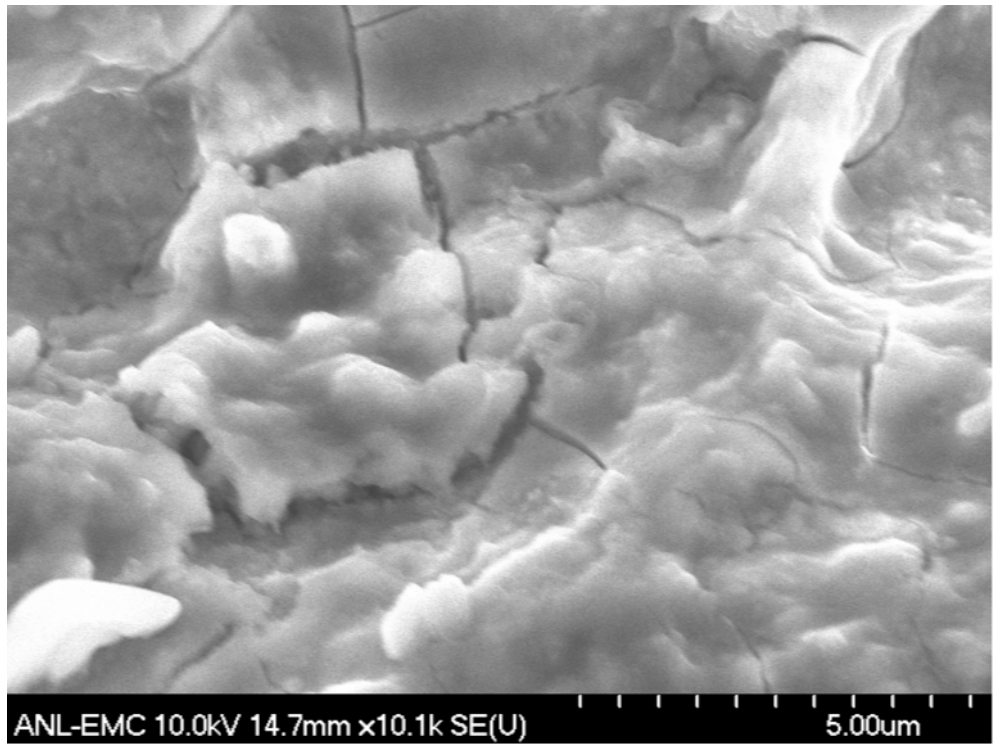}
    \caption{ SEM image of the center of an arc damage crater.  The image shows both cracks, with many crack junctions, and smooth structures characteristic of a chaotic surface smoothed over by surface tension.  There is a wide variety of structures seen in arc damage, and this image is selected to show both cracks and evidence of turbulent structures that have been smoothed by surface tension.  This image should not be considered "typical".}
\end{figure}

During the liquid cooling phase, surface tension would smooth the surface, and the relation between the cooling time and the scale of surface irregularities seen in SEM images can be estimated from the dispersion relation, 
   \[\omega^2 = \sigma |k|^3/ \rho \]
where $\omega, \sigma$ and $\rho$ are the frequency, surface tension constant and density of the liquid metal, and $k$ is the wave number \cite{He,landau}.  For copper structures, where we assume that thin heated volumes sit on essentially cold surfaces, the thermal contraction is approximately 2\% of the dimensions of the melted area.  We find that the typical cooling time constants are in the range of a few hundred ns for accelerator cavities and the structures seen in SEM images of rf cavity damage have radial dimensions on the order of a few microns.   The two stage cooling process seems to result in SEM surfaces that are somewhat smooth at the 1 micron  level, but contain cracks with sharp edges at the 1 - 10 nm level that cover $\sim$2\% of any large solidified area of copper.

\begin{figure}[htb]
    \centering
    \includegraphics*[width=88mm]{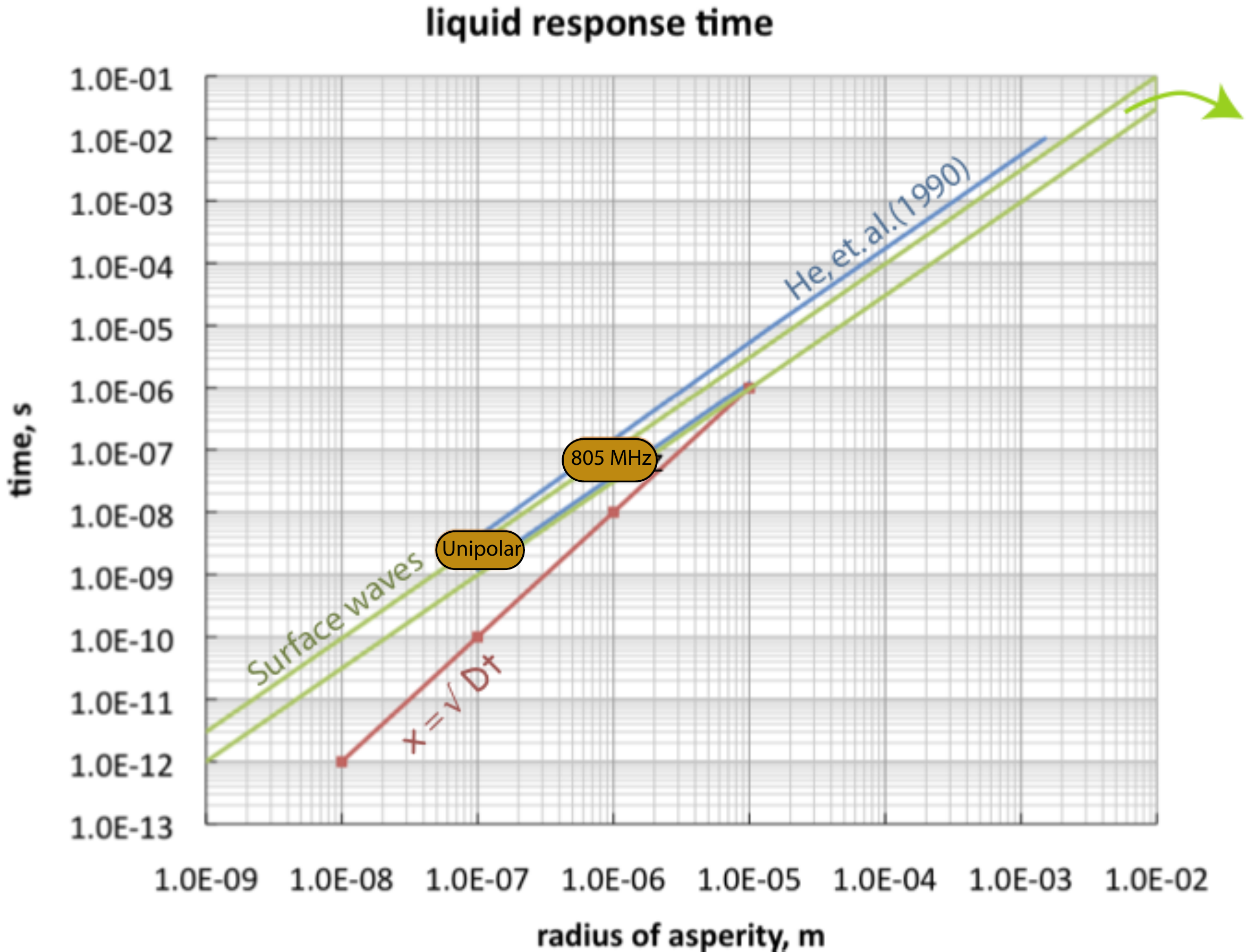}
    \caption{ The relation between the cooling time and structure radius for liquid metals compared with data from 805 MHz cavity arcs and unipolar arcs in coaxial lines.}
\end{figure}

\section{Details}

\subsection{Field Emission}
The process of vacuum breakdown was identified by students of Michaelson and Millikan almost 110 years ago and field emission was of the first mechanisms to be described using quantum mechanics by Fowler and Nordheim in the 1920's when it was discovered that field emission currents were proportional to the applied electric field raised to a high power,  $i \sim E^{n}$, with $n$ around 14 at high surface fields \cite{earhart,FN}.  We find that the standard method of analysis using Fowler-Nordheim (FN) plots to estimate the field enhancement factor can be unnecessarily abstract, and yields a number (the enhancement factor $\beta$) that has little fundamental importance.  We prefer to plot both the experimental data (currents, radiation levels, etc.) against electric field on a log-log plot, along with the FN predictions, although the space charge limit and other experimental parameters can also be displayed.  Using this method, the two lines are offset by factors that can measure the total emitter area, duty cycle, enhancement factor and corrections due to the cavity geometry.  Because field emission currents depend on the electric field raised to a very high power, and enhancement factors are somewhat difficult to measure experimentally, few measurements of the area of field emitters are in the literature.

\begin{figure}[htb]
    \centering
    \includegraphics*[width=70mm]{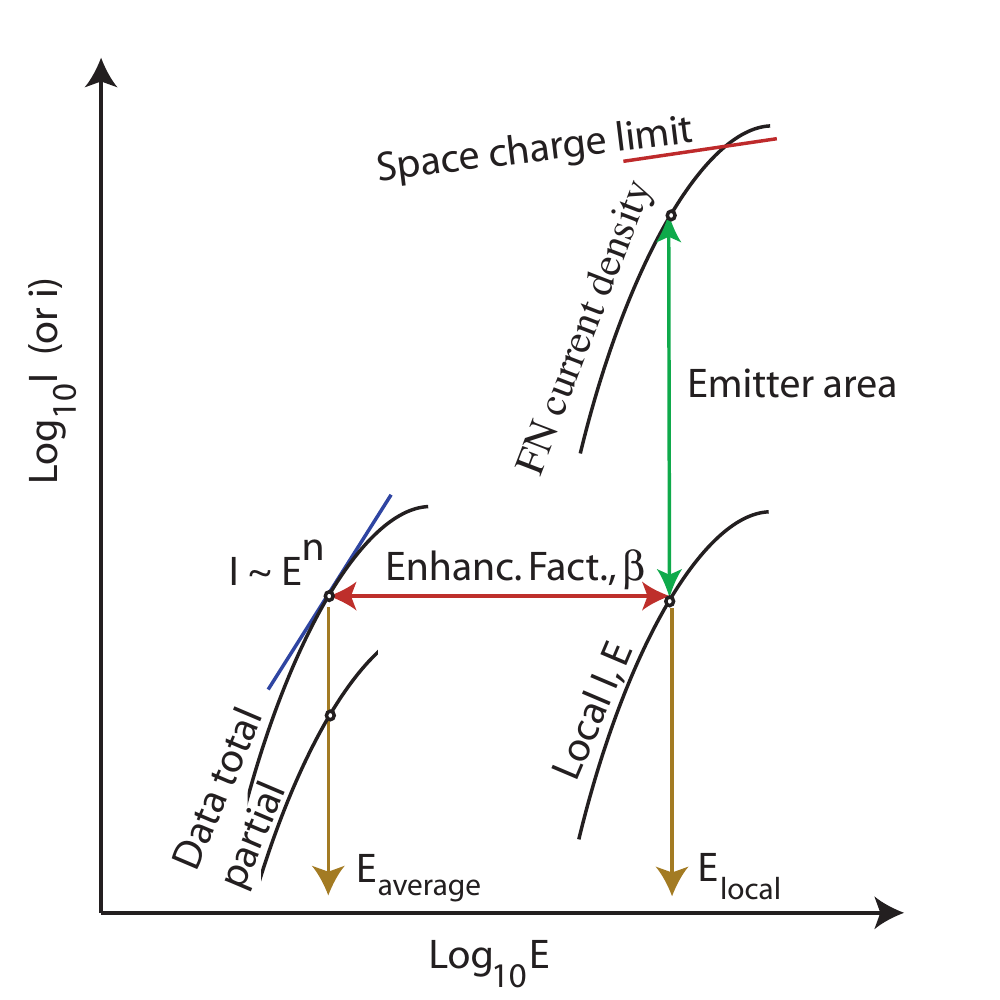}
    \caption{ Field emission can be plotted to display the variables associated with its measurement.  The horizontal and vertical offset of the data and theory curves are essentially the enhancement factor and emitter area but the effects of thermal emission, different work functions, duty cycle, structure geometry etc., as well as systematic and statistical errors in measurement can also be displayed graphically.}
\end{figure}

Following early work by Dyke et. al. showing that Ohmic heating of tungsten needles could produce breakdown, combined with considerable evidence that the required field enhancements and current densities could be produced with cylindrical asperities with rounded ends,  Ohmic heating was widely accepted as an explanation for breakdown, although asperities of the expected dimensions were not found \cite{Dyke}.  
We have shown that cracks, more specifically crack junctions, can provide the required field enhancements, and emitter areas (when many of them are added together) to explain field emission data.

Field emission measurements in rf cavities and high gradient structures have been made and reported in a number of references \cite{PR1}.  These measurements assume that the number of emitters is known and understood.  We assume that a large number of much smaller, localized emitters at crack junctions, contribute as one emitter.

\subsection{Enhancement Factors}
Following Feynman, we describe the surface field of a conductor as a function of the local curvature of the surface, comparing the fields at any two points $a$ and $b$ will give the relation $E_a/E_b = r_b/r_a$, where $r$ is the three dimensional radius \cite{feynman}.  Small radii give high fields.  We find these small radii at crack junctions, where the radii are too small to be resolved by SEM optics.  Numerical analysis has shown that these crack junctions can produce enhancement factors in the range of $\beta \sim 200$.

\begin{figure}[htb]
    \centering
    \includegraphics*[width=70mm]{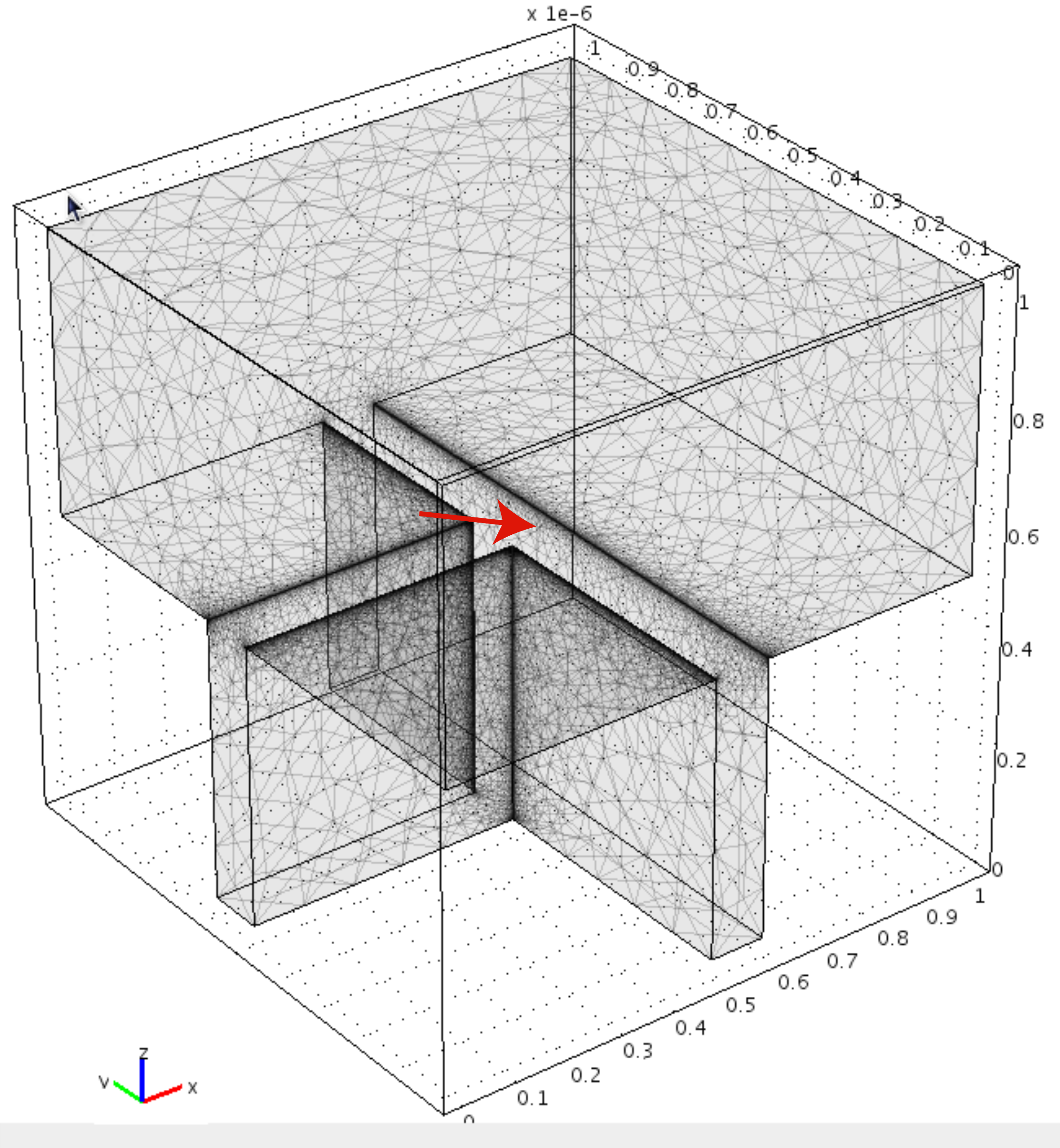}
    \caption{ The crack junction mesh.}
\end{figure}

In another example of field enhancements, one can describe the sheath potential of a tenuous plasma as an enhancement of an applied field that would add the field in the sheath to the externally applied field.  As the plasma density increases, this picture becomes less appropriate.  

\subsection{Breakdown Mechanisms}
It has been shown that breakdown occurs at fields near 10 GV/m.  These local fields would cause mehanical stresses on the order of,
\[ \sigma = \epsilon_0 E^2/2 = 4.4\times 10^8\  \mathrm{MPa} ,\]
which would be pulsed and subject to fatigue,accompanied by high field emission current densities, and perhaps local heating.  Under these conditions mechanical failure would be expecte.  Electrostatic fracture, Ohmic heating, electromigration, fatigue and creep can all explain the the mechanical failures that could trigger breakdown, for example the breakdown rate $BDR \sim E^{30}$ behavior seen in some experiments.  For example, both Ohmic heating and electromigration should be proportional to the current density squared, since field emission produces current densities in the range $j \sim E^{14}$.  Electrostatic fracture is similar to field evaporation, which is governed by processes that produce field scaling from $rate \sim E^{30-150}$, thus seems compatible, although the effects of creep and fatigue are not well understood.  The more difficult problem is to find a mechanism that is compatible with the damage seen in SEM images.

Alltlhough we favor the model of electrical stress and fatigue as a trigger for breakdown, we find that the BD mechanism itself is less interesting than a description of the environments that are highly stressed in many different parameters. It is important to understand nature of asperities and their geometry to understand if it is possible to suppress breakdown.

\subsection{Parameters of Nonideal Plasmas}
Simulations with PIC codes have shown that field emission, significant sheath potentials, high 
densities of neutrals, along with self-sputtering can produce an environment where the density rises essentially exponentially while the electron and ion temperatures remain relatively low.  This increasing density is associated with a decreasing Debye length, 
\[ \lambda_D = \sqrt{\epsilon_0 k_B T / n_e e^2,}  \]
so the number of particles in the Debye sphere eventually becomes less than one and the nonlinearity parameter, $\Theta$ ,which measures the ratio of the electrostatic potential energy divided by the kinetic energy of the plasma, becomes large \cite{morozov}.  

Recent numerical analysis of high density, non-ideal plasma sheaths has shown that for high density plasmas the properties of the plasma can be estimated using molecular dynamics.   The corrections to simple estimates of sheath potential, Debye length and surface electric field required by the non-ideality condition due to the high densities involved are not large.  Plasma densities were estimated from the scale of damage, where turbulence of produced by the plasma pressure is balanced against the smoothing produced by the surface tension.  Since it is difficult to know the cooling time with much precision, these measurements function as an upper limit on the scale of turbulence and a lower limit on the plasma density.   This procedure produces estimates of the surface plasma density $n \sim 10^{25}$ m$^{-3}$, surface electric fields $E \sim 2 \times 10^9$ GV/m,  for electron temperatures of 10 eV \cite{morozov}.

\subsection{Space Charge Oscillations}
PIC codes have shown that when field emitters can ionize dense gas near the surface, a positively charged  plasma is produced, and the sheath potential of the plasma that is created increases the field on the field emitters until they become space charge limited.

The space charge limit for continuous currents between two plates is expressed using the Child-Langmuir Law,
\[ I =  \frac{4 \epsilon_0}{9}\sqrt{2e/m_e}  \frac{SV^{3/2}}{d^2} , \]
where $I$ is the anode current,  the current density, and S the anode surface inner area \cite{gray}. 
While the Child-Langmuir Law applies to thermionic emission, the application of this idea to field emission is not entirely straightforward. Thermionic emission of electrons is essentially constant, with fluctuations governed by variations in the temperature of the emitter.  With field emission, however, the current density is proportional to the electric field to some high power, ($i \sim E^{14}$), thus, fluctuations in the electric field will instantly alter the emitted current density, and fluctuations in the density of emitted electrons will immediately alter the electric field.  These processes can produce fluctuations.  

PIC code results show that the space charge limited current is not continuous on a microscopic scale.  Electrons are emitted from the surface in bunches, they move a few microns away from the cathode where they produce a negatively charged electron cloud that erodes due to electrons moving both toward and away from the cathode.  This repetitive behavior produces an oscillation in the field emitted current at a frequency of about 1 THz \cite{movies}.  We are not aware of experimental observation of this phenomenon.  

\subsection{Plasma Fluctuations and Quenching}
Vacuum arcs can be unstable.  Fluctuations in the optical emission of arcs have been recorded in streak camera experiments and one of the defining properties of unipolar arcs is their random, discontinuous, trail of surface damage.  Optical fluctuations occur at frequencies up to a few hundred MHz.  We describe the fluctuations seen in unipolar arcs as a similar mechanism to the fluctuations in the space charge limited field emission described above in Section IIIA.  

Although over long time scales the plasma should maintain quasi-neutrality, the mechanisms controlling the electron and ion densities are quite different, and have different timescales.   Since the field emitted current density will be proportional to, $i_{FE} \sim E^{16}$, the field emission current will respond instantly and nonlinearly to changes in the surface field, and the electrons can be rapidly thermalized in a dense plasma.  

We assume that the fundamental ion density increase is governed by self-sustained self-sputtering,
\[ \alpha \beta \gamma >1  ,\]            
where $\alpha$  is the probability that a sputtered cathode atom becomes ionized, $\beta$ is the probability that the ionized atom returns to the cathode, and $\gamma$ is the sputtering yield \cite{Anders}.  The ion density, $n_i$, should respond slowly, since the time constant for density changes would depend on collisional diffusion \cite{rose} in the arc,
\[  \partial n_i/\partial t = D\  \nabla^2 n_i    ,\]
which is  a function of the density, $n_i$, since the diffusion constant is inversely proportional to the plasma density,
\[  D = v_{th}/ 3\nu \sim 1/n_i   .\]  
As the arc evolves and the density increases, the large, dense arcs should become more stable to ion density fluctuations, with time constants proportional to, $\tau_i \sim n_i $, the time constant for field emission, however, should not change and the electron thermalization time should become shorter as the density increases like, $\tau_e \sim 1/n_i$.  This difference between the ion and electron density stability could complicate the ability of the plasma to maintain quasi-neutrality under rapid high current field emission.

As the arc evolves, surface fields created by the sheath potential become large enough to produce field emission currents that can short out the sheath potential and locally quench the arc before quasi-neutrality can be established.  As shown in reference \cite{morozov}, the current required to short out a plasma sheath in time $\Delta t$ is equal to,
  \[ i_s = \epsilon_0 E/\Delta t  ,\]
for times of $\Delta t =$ 1 ns, the required current would be $\sim$30 MA/m, which could be produced by a field of $\sim$3 GV/m, which is compatible with simulations produced by both PIC and MD codes.  The remaining dense plasma is then either able to restart the arc nearby, or, after a time required to equilibrate the locally dense plasma, restart in the same location.  These densities are compatible with data taken with 805 MHz rf structures, which have arc damage diameters of 0.5 mm diameter and shorting currents on the order of 10 A.  This argument seems to preclude plasma / surface fields significantly larger than 3 GV/m and current densities larger than 30 MV/m.  

The comparatively low current density of 30 MA/m$^2$ is not large enough to produce significant Ohmic heating of the surface. This seems to conflict with the current densities required by the ecton model of G. Mesyats  \cite{mesyats}.  In that model, current densities of $\sim10^{13}$ A/m$^2$ are required to produce a local Ohmic heating explosion of the liquid metal that maintains the arc.  

\subsection{Frequency Dependence of Gradient Limits}
The maximum operating gradient of a given rf or DC system could operate should be a function of two variables; 1) the maximum local field at which the surface would fail, due to tensile stresses, heating, electromigration, fatigue or some other effect, and, 2) the overall design of the system itself, which determines the stored energy deposited through the arc, the way power is applied, discharge length, the way the power to the arc is turned off (suddenly or slowly) which all seem capable of affecting the surface damage, and ultimately the field enhancements seen by the surface  \cite{hassanein}.
 
A large body of data showed very early that DC breakdown occurred when local fields reached 7 - 10 GV/m over many orders of magnitude variations in the gap length.  Although there are not many rf measurements, data also show that this threshold also seems to apply to systems around 1 GHz.  These results imply that there is no frequency dependence to high gradient breakdown as a function of the local electric field, $E_{local}$.  On the other hand, there is an extensive literature that show that higher frequencies achieve higher gradients.

\subsection{Magnetic Fields}
We have found that under some conditions the maximum rf electric field that can be maintained in the presence of an externally applied electric field is reduced from that seen without the external magnetic field. 

The beam optics of field emitted beams in magnetic fields have been studied experimentally.  The beams were actually found to be hollow, with a radius that was directly proportional to the applied electric field and inversely proportional to the square of the static magnetic field.  This is consistent with a picture of field emission from a foil-less diode, where the electric and magnetic fields were not parallel.

SEM images of arc damage in copper with a magnetic field shows that surface cracking is confined to a very small area (10 - 20) $\mu$m in diameter surrounded by a much larger area that shows signs of being melted.  We assume that the radial growth of the arcs were confined by the magnetic field, and when the arcs cooled they cooled from the outside in, leaving the last metal to solidify to absorb all the thermal contraction.  These central damaged areas have a much higher crack density, that we associate with the lower electric fields that could be maintained on the surface.

\subsection{Other Environments}
Although we consider arcing primarily in the context of rf linacs, It is useful to see how generally these arguments apply to other environments, such as tokamak first walls \cite{ITER}, laser ablation \cite{kajita}, micrometeorite impacts \cite{meteorites} and possibly such examples as electron beam welding.  Arc damage in laser ablation, tokamak first walls and micrometeorite impacts seems essentially identical.

The study of unipolar arcs is not an active field.  Although vacuum arcing phenomena are seen in many environments, there is little contact and little coordination between different approaches used in these fields.   Arc damage seems to be quite similar between micrometeorite impacts, laser ablation targets and tokamak first walls, however the lack of a model or a common approach to understanding the basic mechanisms at work in unipolar arcs has further slowed progress.

\section{Useful Experiments}
There have been 110 years of experimentation on vacuum arcs, most of them guided, to some extent, by modeling and theory, nevertheless there is still disagreement about the nature of these arcs and the mechanisms that drive them.  We believe the reason for this situation is that the arcs are small and unpredictable, and many parameters (which are individually hard to measure) evolve very rapidly over a many orders of magnitude. While models exist, theory and modeling are complicated by the large number of mechanisms that seem to be involved in arc evolution and high density plasmas, that require a complicated, non-Debye analysis of even basic properties.

There are a number experimental directions that could prove promising.  Space charge oscillations have not been seen  as far as the authors are aware, although they would be somewhat difficult to detect because of the high frequency involved.  Likewise, quench of unipolar arcs has never been studied in any detail, as it is difficult to access the surface underneath the plasma.  And lastly, it would be useful to have more systematic data on the damage mechanism in normal arcs.  While this field has produced considerable data, the problem is that little of it was systematically selected and documented.

\section{Conclusions}
We have demonstrated that the physics of unipolar arcs and surface cracking seems highly relevant to the the phenomenon of arcing.   Crack junctions can provide the high field enhancements seen in experimental studies of field emission and breakdown, they are formed naturally as arc damage cools and they are capable of triggering breakdown events when they fracture.  Likewise, unipolar arcs are the prototypes of single sided arcs, that can function as cathode spots.  We find that the physics of these objects, which has not received much specific attention is relevant to many fields.

Although unipolar arcs have been called ubiquitous, the literature on this phenomenon, both experimental an theoretical, is not extensive.  We believe one reason for this is that the dense plasmas and plasma / surface interactions require very specific techniques to cope with the nonideal (non-Debye) plasmas, that are not well advanced.

We have shown that cracks are commonly seen in SEM images of arc damage and described how they are produced as the liquid metal cools below the melting point to room temperature. We have shown that cracks can produce the required field enhancements to explain field emission data data and can produce fractures that would trigger breakdown events.  Although unipolar arcs, and non-ideal plasma surface interactions are not well studied, we have shown that field emission of electrons produced in the plasma sheath can produce currents (1 - 1000 A) sufficient to short out rf structures, should cause the sort of damage seen in SEM images. These plasmas should be unstable and possibly self-quenching as seen in optical fluctuations and surface damage in a variety of experiments.   

Although the internal structure and evolution of arcs and arcing has not evolved to any unanimity in the field, we find that the physics of cracks and unipolar arcs seem highly relevant and perhaps fundamental to these phenomena.

\section{Acknowledgements}
We  thank the staff of the Accelerator and Technical Divisions at Fermilab and the Muon Accelerator Program (MAP) for supporting and maintaining the MAP experimental program in the MTA experimental area.    The work at Argonne is supported by the U.S. Department of Energy Office of High Energy Physics  under Contract No. DE-AC02-06CH11357.

\newpage

The submitted manuscript has been created by UChicago Argonne, LLC, Operator of Argonne National Laboratory ("Argonne").  Argonne, a U.S. Department of Energy Office of Science laboratory, is operated under Contract No. DE-AC02-06CH11357.  The U.S. Government retains for itself, and others acting on its behalf, a paid-up nonexclusive, irrevocable worldwide license in said article to reproduce, prepare derivative works, distribute copies to the public, and perform publicly and display publicly, by or on behalf of the Government.

\end{document}